\documentclass[runningheads]{llncs}

\usepackage{eccv}

\usepackage{eccvabbrv}

\usepackage{graphicx}
\usepackage{booktabs}
\usepackage[linesnumbered,ruled]{algorithm2e}
\usepackage{authblk}

\usepackage[accsupp]{axessibility}  

\usepackage{hyperref}

\usepackage{orcidlink}

\begin{document}

\title{WindPoly: Polygonal Mesh Reconstruction via Winding Numbers} 

\titlerunning{WindPoly}

\author{Xin He\inst{}\orcidlink{0009-0005-6204-8557} \and
Chenlei Lv\inst{}\orcidlink{0000-0002-8203-3118} \and
Pengdi Huang\inst{}\orcidlink{0000-0002-9926-4393} \and
Hui Huang\thanks{Corresponding author}\inst{}\orcidlink{0000-0003-3212-0544}}

\authorrunning{X. He, C. Lv, P. Huang, and H. Huang}

\institute{Visual Computing Research Center, Shenzhen University \\
\email{hhzhiyan@gmail.com}}

\maketitle

\begin{abstract}
  Polygonal mesh reconstruction of a raw point cloud is a valuable topic in the field of computer graphics and 3D vision. Especially to 3D architectural models, polygonal mesh provides concise expressions for fundamental geometric structures while effectively reducing data volume. However, there are some limitations of traditional reconstruction methods: normal vector dependency, noisy points and defective parts sensitivity, and internal geometric structure lost, which reduce the practicality in real scene. In this paper, we propose a robust and efficient polygonal mesh reconstruction method to address the issues in architectural point cloud reconstruction task. It is an iterative adaptation process to detect planar shapes from scattered points. The initial structural polygonal mesh can be established in the constructed convex polyhedral space without assistance of normal vectors. Then, we develop an efficient polygon-based winding number strategy to orient polygonal mesh with global consistency. The significant advantage of our method is to provide a structural reconstruction for architectural point clouds and avoid point-based normal vector analysis. It effectively improves the robustness to noisy points and defective parts. More geometric details can be preserved in the reconstructed polygonal mesh. Experimental results show that our method can reconstruct concise, oriented and faithfully polygonal mesh that are better than results of state-of-the-art methods. More results and details can be found on https://vcc.tech/research/2024/WindPoly.
\end{abstract}

\section{Introduction}
\label{sec:intro}

As an efficient representation of a three-dimensional object, a polygonal mesh has significant advantages in the shape reconstruction task. It carries structured face information with geometric consistency while significantly reducing data volume. Especially for architectural point cloud representation, a polygonal mesh provides accurate and concise geometric structures that are useful for visualization and feature analysis in urban scenes.

The mainstream reconstruction technical routes include two-step mesh reconstruction and low-poly meshing directly. For the first route, the related methods attempt to reconstruct complex mesh from point cloud to keep more geometric details~\cite{hou2022iterative,wang2022restricted,xu2022rfeps,kazhdan2013screened,kazhdan2006poisson,bolitho2009parallel}. Then, they employ simplification strategies~\cite{salinas2015structure, lescoat2020spectral} to achieve concise meshes for data compression. However, some important geometric features are broken by the simplification to a certain extent. In addition, the performance of the reconstruction is inevitably reduced for incomplete point clouds.  For the second route, the methods ~\cite{bauchet2020kinetic,li2016manhattan,nan2017polyfit,lafarge2013surface} decompose the 3D space for the incomplete point cloud and directly establish a polygonal mesh to provide a concise structural representation. But they are sensitive to outliers of raw point cloud and lose some accurate geometric features.

In this paper, we propose a robust polygonal mesh reconstruction method to implement low-poly meshing for architectural point clouds. It includes three core components. Firstly, we detect polygonal planes from the raw point cloud, which is inspired by assembling-based surface reconstruction~\cite{nan2017polyfit}. Secondly, we design an adaptive spatial partitioning scheme that iteratively segments the internal space represented by achieved planes. It is used to control the scope of planar intersection. A set of convex polyhedrons is constructed to fill the internal space. With the polygon-based winding numbers optimization, the polygonal mesh is constructed from the convex polyhedrons, which captures concise representation and more geometric details from noisy and unorganized point clouds. The contributions can be summarized as:
\begin{itemize}

\item We present a polygonal plane detection method that clusters points into regular planes without normal vector analysis. It is robust to outliers and noisy points while providing a structural geometric representation for the incomplete and unorganized point cloud.

\item We design an adaptive spatial partition to establish a set of convex polyhedrons, which are used to represent the internal space of the architectural point cloud. A large number of plane-based intersection calculations are avoided which improve efficiency and enhance the accuracy of internal geometric details.

\item We propose a polygon-based winding numbers optimization to achieve the final polygonal mesh. The optimization strategy fully inherits the advantages of the winding number for surface orientation. At the same time, the strategy considers the consistency constraint with the original input point cloud. It further enhances the quality of the reconstructed polygonal mesh.

\end{itemize}
\section{Related Work}\label{sec:Related Work}
Related methods for polygonal mesh reconstruction can be concluded into three parts: mesh reconstruction, mesh simplification, and low-polygon meshing.

\subsection{Mesh Reconstruction}
Basically, the mesh reconstruction can be divided into two categories: explicit surface reconstruction and implicit one. The explicit surface reconstruction is to establish continuous representations for discrete forms by establishing explicit local neighborhood. Representative solutions include ball pivoting~\cite{bernardini1999ball}, scale space~\cite{Digne2011Scale}, Delaunay triangulation~\cite{cohen2004greedy}, and voxel-based reconstruction~\cite{wang2005reconstructing}. Following the development of deep learning, many researchers attempt to learn the prior knowledge from modeling experiences to guide 3D reconstruction~\cite{williams2019deep, hanocka2019meshcnn, hanocka2020point2mesh, chen2020bsp, yang2021unsupervised, huang2022neural}. Compared to the explicit surface reconstruction, the implicit one solves an implicit function, which defines the surface as the zero-level set of the function. Some representative methods include marching cubes~\cite{lorensen1987marching, nielson2003marching}, Poisson surface reconstruction~\cite{kazhdan2006poisson, kazhdan2013screened, kazhdan2020poisson, hou2022iterative}, and radial basis function~\cite{carr2001reconstruction, morse2005interpolating, xu2022hrbf}. They are more flexible and can handle topologically complex shapes.

\subsection{Mesh Simplification}
To preserve more geometric features, reconstructed meshes usually carry more vertices and faces which increase the computational cost of mesh-based rendering and analyzing. Therefore, mesh simplification schemes are proposed to compress the 3D model. Such schemes include geometric-based approximation~\cite{cohen2004variational, calderon2017bounding, lescoat2020spectral}, Delaunay-based remeshing~\cite{yi2018delaunay, rakotosaona2021differentiable}, structural simplification~\cite{salinas2015structure, gao2022low}, hierarchy strategy~\cite{li2021feature}, and intrinsic analysis~\cite{lv2022adaptively, Derek2023Surface}. These schemes can approximate a polygonal mesh or reconnect a simplified mesh from the original complex one while keeping some important geometric features. However, these methods require high-precision mesh input, which makes the processing flow cumbersome and unstable.

\subsection{Low-Poly Meshing}
To address the limitation of mesh simplification, low-poly meshing is proposed to directly establish concise polygonal meshes based on raw inputs. Most representative methods utilize the idea of structural reconstruction to implement the meshing process, including building blocks~\cite{mehra2009abstraction, li2016manhattan}, structure-preserving~\cite{lafarge2013surface}, surface elements~\cite{kelly2017bigsur} and plane hypothesis~\cite{nan2017polyfit, fang2018planar, bauchet2020kinetic, fang2020connect}. These methods extract basic representation elements from raw point clouds to construct polygonal meshes while approximating objects from incomplete geometric structures. However, there are limited by normal dependency and lower computational efficiency. 
\section{Methodology}\label{sec:method}

\begin{figure}[t]
	\centering
    \includegraphics[width=\linewidth]{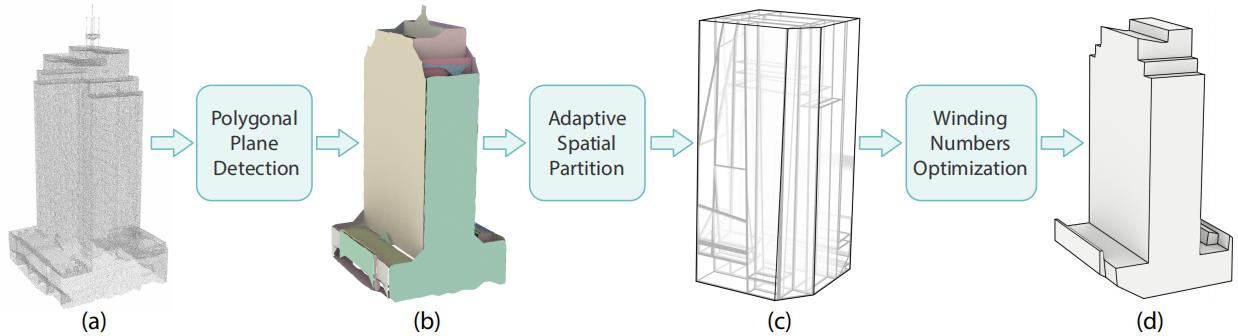} 
	\caption{Pipeline of WindPoly. (a) raw point cloud, (b) detected polygonal planes, (c) adaptive spatial partition, (d) output by polygon-based winding numbers optimization.}
	\label{feP}
\end{figure}

\textbf{Overview.} The proposed WindPoly can be regarded as a low-poly remeshing scheme. It can be concluded as three parts: polygonal plane detection, adaptive spatial partitioning for convex polyhedron generation, and polygon-based winding numbers optimization. The polygonal plane detection is used to detect primitive planes from the raw point cloud in order to achieve initial structure information. Then, the convex polyhedron generation implements internal structure fitting by using adaptive spatial partitioning. This results in a set of coarse polyhedral elements without correct directions. To achieve the final polygonal mesh, the polygon-based winding numbers optimization is implemented to orient polyhedral elements with global consistency. Some ambiguous geometric structures are improved during the orientation and consequently obtain more precise internal geometric details. The pipeline is shown in Fig. \ref{feP}. In the following parts, we explain the implementation details.

\subsection{Polygonal Plane Detection}

The scanned raw point cloud inevitably carries noisy points and outliers with random distributions. Therefore, we employ a pre-processing step to increase the robustness for raw point clouds. Firstly, we simplify the raw point cloud by Poisson resampling. It reduces the scale of the point cloud and optimizes its densities to be uniform. Then, we remove the outliers based on an automatic neighbor analysis program that is similar to the solution in \cite{lv2022intrinsic}. Let $p_i$ represent a point in a point cloud $P$, $N(p_i)$ is the neighbor set of $p_i$ which is achieved by KNN. The judgement of outlier set $\{p_o\}$ can be formulated as
\begin{equation}
\{p_o\} = \{p_i|p_i\not\in N(p_j), p_j\in N(p_i), p_i,p_j\in P\},
\end{equation}
where $p_j$ is the neighbor point of $p_i$, and $p_i$ is judged as an outlier when it does not belong to neighbor sets of its neighbors. The formulation is established based on the manifold compactness.

Based on the pre-processed point cloud, we implement polygonal plane detection for raw point clouds. The usual approach is to estimate the normal vector of each points and cluster them to fit related plane~\cite{rabbani2006segmentation}. It is a normal vector dependency scheme with limited robustness. To implement a scheme with normal vector independence, we present a practical solution based on existing methods. We firstly extract candidate planes by fitting planar primitives (FPP)~\cite{yu2022finding}. It extracts basic planes based on primitive configuration achieved by
\begin{equation}
U(x)=\omega_fU_f(x)+\omega_sU_s(x)+\omega_cU_c(x),
\end{equation}
where $U_f$, $U_s$, and $U_c$ represent fidelity, simplicity and completeness energies with related weights $\omega_f$, $\omega_s$, and $\omega_c$, which describe the clustering property in local neighborhoods. Then, we check the planes and combine them according to the plane refinement regulation (PRR) mentioned in~\cite{nan2017polyfit}. It can be represented as
\begin{equation}
\begin{array}{c}
acos\left\langle n_i,n_j\right\rangle<\theta_{the},\\
N_t=min(\vert F_i\vert,\vert F_j\vert)/5,
\end{array}
\label{e3}
\end{equation}

where $n_i$ and $n_j$ are normal vectors of planes $F_i$ and $F_j$, $|F_i|$ and  $|F_j|$ are points belonging to the two planes, $\theta_{the}$ ($10^{\circ}$ by default) and $N_t$ are control parameters. Once the common point number between two planes is larger than $N_t$ and simultaneously satisfies the first condition in Eq.~\eqref{e3}, the plane-based combination is triggered, the two related planes are merged.
The required planes are achieved until all candidate planes are checked and combined. The implementation of the method can be concluded in Algorithm~\ref{A1_Plane}.

\IncMargin{1em}
\begin{algorithm}[t]
  \SetKwInOut{Input}{Input}
  \SetKwInOut{Output}{Output}
  \SetKwInOut{Parameter}{Param.}
    \caption{Polygonal Plane Detection}  
    \label{A1_Plane}
    \Indentp{-1em}
    \Input{Pre-processed point cloud $P$}
    \Output{Polygonal plane set $\{F\}$}    
    \Indentp{1.3 em}
    Initialization: $\{F\}_{c} \gets \textsc{FPP}(P)$ \\    
    \ForEach {plane $F_i \in \{F\}_{c}$}
    {
        \ForEach {plane $F_j \in \{F\}_{c}, j\neq i$}{
          \If {\textsc{PRR}($F_i, F_j$)}{
              Combine $F_i$ into $F_j$\\
              Update $\{F\}_{c}$\\
          }
        }              
    }    
    $\{F\} \gets \{F\}_{c}$
\end{algorithm}
\DecMargin{1em}

\subsection{Adaptive Spatial Partition}

Based on detected polygonal planes, we propose an adaptive spatial partition to generate convex polyhedrons for internal structure perception. It inherits the idea of iteratively generating polyhedrons in the convex polyhedral space while using a concise strategy to fit the internal structure without normal vector assistance. The method includes two basic components: convex polyhedral space construction and iterative convex polyhedron searching. 

\begin{figure}[t]
  \centering
  \includegraphics[width=.95\linewidth]{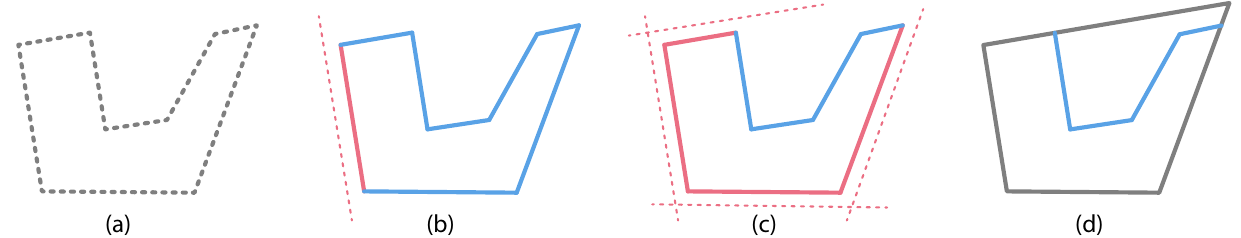} 
  \caption{External plane detection in 2D vision. Gray dotted lines (a) represent the raw point cloud. Red lines represent the external planes (b), blue lines are the other planes located in same side of the external planes (c). Finally, the convex polyhedral space represented by external planes is achieved (d).}  
  \label{f3_convex}
\end{figure}

The convex polyhedral space is the convex hull of the object, which is constructed by the external planes selected from the primitive elements. The external plane is detected based on the regulation that is all other primitive elements should be located in the same side of the plane. An instance is shown in Fig. \ref{f3_convex}. It should be noticed that some outliers and inaccurate planes take influences for external plane judgement. To improve the robustness, we implement an additional check for the intersection of two planes. Let plane $F_i$ to be a cross plane of $F_0$, we compute the $\alpha$ shape to achieve the boundary point set $\{p\}_b^0$ of $F_0$. The $\alpha$ shape represents a point-based region that describes the enveloping shape of associated points. We check points in $\{p\}_b^0$ and delete ones that satisfy $dist(p_j^0,F_i)<\sigma$, where $dist$ is the distance between $p_j^0$ and $F_i$. If all the remaining boundary points of $\{p\}_b^0$ are located on one side of $F_i$, then the entire plane $F_0$ is judged to be on the side of $F_i$. The additional check improves the practicality of external plane detection.

\begin{figure}[t]
  \centering
  \includegraphics[width=.94\linewidth]{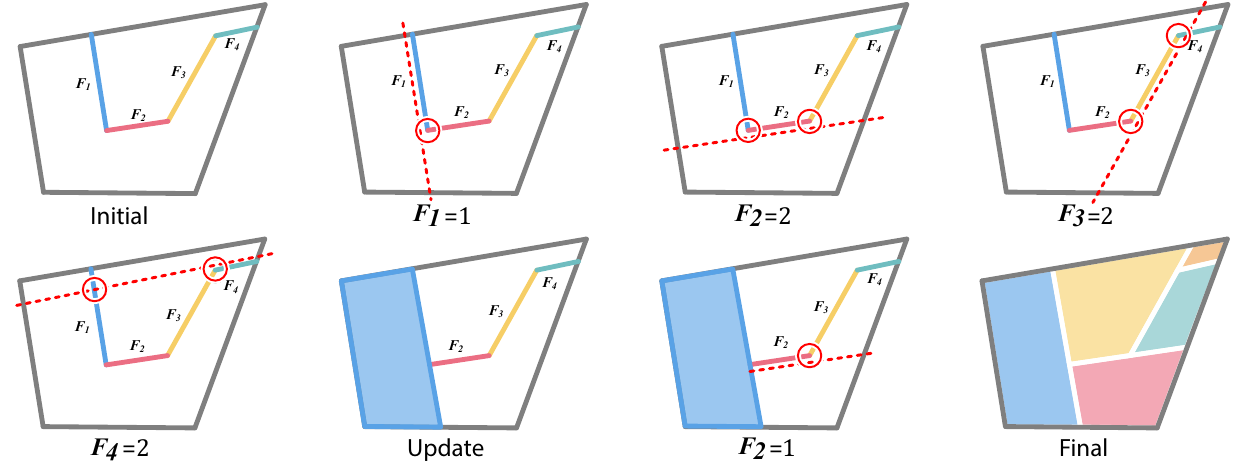} 
  \caption{An instance of adaptive spatial partition in 2D vision. In the initial step, intersection numbers of related planes $F_1\sim F_4$ are 1, 2, 2, 2. After partition according to the regulation, the convex polyhedral space is divided into a set of convex polyhedrons.}  
  \label{f4_class}
\end{figure} 

In the convex polyhedral space, we establish convex polyhedrons to perceive the internal structure. PolyFit~\cite{nan2017polyfit} has developed a solution that implements exhaustive partition to generate convex polyhedrons. However, it requires redundant intersection calculations for all planes with poor robustness. To solve the problem, we employ the adaptive spatial partition to segment the convex polyhedral space. Each internal plane is expanded and intersects with its spatial sub-region. Benefited from the adaptive spatial partition, the efficiency of convex polyhedron generation can be significantly improved. The adaptive spatial partition can be concluded as an iterative scheme. 

\begin{figure}[t]
  \centering
  \includegraphics[width=.95\linewidth]{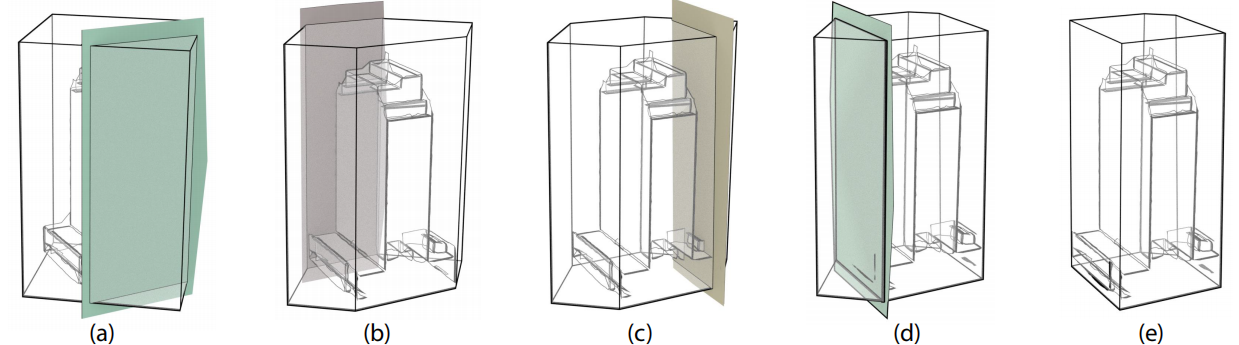} 
  \caption{An instance of adaptive spatial partition in 3D vision. According to the separating plane searching (a)-(d), the 3D structure can be achieved with internal geometric details (e).} 
  \label{f4_cut}
\end{figure}

As an initial step, the internal planes are collected into the set $\{F\}_{int}$. Intersection numbers $\{num\}_{int}$ between each internal plane and other internal ones are computed and stored at the same time. Then, we select the separating plane $F_i$ with minimum intersection number of $\{num\}_{int}$. According to the $F_i$, the original convex polyhedral space represented by external planes are divided into two parts. We iteratively search for separating planes in the corresponding part and continuously segment the subspaces until all planes are checked. Each convex polyhedral subspace corresponds to a convex polyhedron. Finally, a set of convex polyhedrons can be constructed. The intersection calculations between internal planes are controlled in the related subspace. It avoids redundant calculations while improving accuracy. The adaptive spatial partition is concluded in Algorithm~\ref{A3_partition}.

\IncMargin{1em}
\begin{algorithm}[t]
  \SetKwInOut{Input}{Input}
  \SetKwInOut{Output}{Output}
  \SetKwInOut{Parameter}{Param.}
    \caption{Adaptive Spatial Partition}  
    \label{A3_partition}
    \SetKwFunction{Fun}{AdaptiveFun}
    \Indentp{-1em}
    \Input{Internal plane set $\{F\}_{int}$}  
    \Output{Convex polyhedron set $\{CP\}_c$}
    \Indentp{1.3 em}    
    Initialization\\
    $\{num\}_{int} \gets$ intersection numbers of $\{F\}_{int}$\\  
    \Fun{$\{F\}_{int}$, $\{num\}_{int}$}:{\\
    Search the $F_i$ from $\{F\}_{int}$ with minimum value from $\{num\}_{int}$\\
    Cut convex polyhedral space into two sub-spaces\\
    \If {sub-space$_{1,2}$ has planes}{
           $\{F\}_{s1,2}\gets \{F\}_{int}$\\
           \Fun{$\{F\}_{s1,2}$, $\{num\}_{int}$}    
    }\ElseIf{}{
           $\{CP\}_c\gets$ sub-space$_{1,2}$    
    }
    \KwRet\;
    }   
\end{algorithm}
\DecMargin{1em}

For the separating plane searching, if there are multiple planes with same minimum value, we select the one with the largest area. In Fig.~\ref{f4_class}, we show an instance for the adaptive spatial partition. The partition process tends to search the subspaces from the outside to the inside in the convex polyhedral space (Fig.~\ref{f4_cut}). After that, a coarse structural representation based on a set of polyhedrons is achieved which has detected the internal regions. 

\subsection{Polygon-based Winding Numbers}
The convex polyhedrons take redundant faces for polygonal mesh representation, which should be removed for the final polygonal mesh representation. Based on the generated convex polyhedrons, we introduce the third part that is to employ winding numbers optimization to determine the orientation of the candidate faces and decide whether to remove the related polyhedrons. As a mature orientation strategy, winding numbers are widely used in recent works~\cite{xu2023globally,feng20023winding}, which optimize a scalar field to define the inside and outside of a closed surface. Generally, such optimization is processed on redundant points which requires a huge computation cost. In our framework, we propose a polygon-based winding number strategy that orients faces directly. Fewer points are used for the computation that significantly improve the efficiency. It can be represented as
\begin{equation}
w(q)=\sum_{i=1}^Na_i\frac{(p_i-q)\cdot n_i}{4\pi\left\|p_i-q\right\|^3},
\label{e4}
\end{equation}
where $q$ is a polyhedron centroid, $p_i$ is the center of face $F_i$ of related polyhedron, $a_i$ is the area of $F_i$, and $n_i$ is the face-based normal vector. Then, we complete the redefinition of parameters for winding numbers. The value of $w(q)$ can be computed which represents the inside or outside direction. The numerical distribution of winding numbers corresponds to the global consistency of normal vectors, which is suitable for orientation.

\begin{figure}[t]
  \centering
  \includegraphics[width=\linewidth]{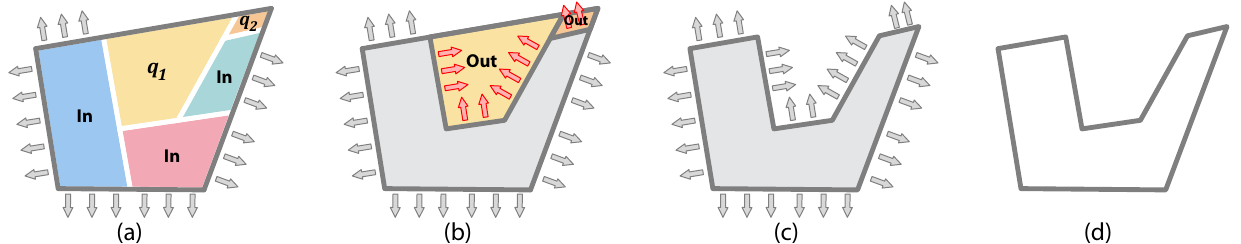}
  \caption{An instance of polygon-based winding numbers optimization. According to the binary labels of related centroids, the polygonal mesh can be extracted from the convex polyhedron set.}  
  \label{f5_winpoly}
\end{figure} 

Let $\{CP\}_c$ represent the convex polyhedron set, we collect candidate faces $\{F\}_{cp}$ from $\{CP\}_c$, which meet the condition $A(F_i)_{\alpha}/A(F_i)>T_r$, ($A(F_i)_{\alpha}$ is the $\alpha$ shape area of $F_i$, $A(F_i)$ is the area of $F_i$, $T_r$ is the control threshold). Then we search the outside faces $\{F\}_{out}$ from candidate faces based on the convex hull of the model, and assign related normal vectors $\{N\}_{out}$ to the outside of the convex hull. A group of faces with normal vectors for winding number computing has been obtained as initialization ($p_i\in\{F\}_{out}, n_i\in\{N\}_{out}$ for Eq.~\eqref{e4}). Next, we define a judgement energy to determine whether a polyhedron should be removed, which can be represented as
\begin{equation}
E_{dir}=W(\{q\})+V(\{q\}),
\label{e4_1}
\end{equation}
where $q$ represents the polyhedron centroid mentioned in Eq.~\eqref{e4}. The purpose of Eq.~\eqref{e4_1} is to achieve a set of labels of $\{q\}$ by minimizing the direction-based energy $E_{dir}$. Once the label of the polyhedron centroid is provided, the judgement of the related convex polyhedron can be processed. The optimization consists of two terms, $W(\{q\})$ and $V(\{q\})$. The first term $W(\{q\})$ employs the winding numbers optimization to compute the face-based direction energy. It provides a geometrically consistent optimization method for determining the optimal orientation combination of polyhedrons, which can be represented as
\begin{equation}
W(\{q\})=\sum_{q_k\in\{q\}}\widetilde{w}(q_k),
\label{e5}
\end{equation}
\begin{equation}
\widetilde w(q_k)=\left\{\begin{array}{c}1-w(q_k),q_k\in{\{q\}}_{in}\\w(q_k),q_k\in{\{q\}}_{out}\end{array}\right.,
\label{e6}
\end{equation}
where $q_k$ is a polyhedron centroid, $q_k\in\{q\}$, $\widetilde{w}(q_k)$ is used to normalize winding number $w(q_k)$ for optimization, $\{q\}_{in}$ and $\{q\}_{out}$ are subsets of $\{q\}$ with related binary labels, which determine whether the polyhedron should be retained.
Once the label of $q_k$ is decided, the first term of direction energy can be achieved. The second term is used to check fitting degree between faces and original point cloud, which assists the determination of direction judgement. It can be computed by
\begin{equation}
V(\{q\})=\sum_{F_l\in\{q_i, q_j\}}(1-A_\alpha(F_l)/A(F_l)),
\label{e7}
\end{equation}
where $F_l$ is a common face between the adjacent convex polyhedrons $q_i$ and $q_j$, $A_\alpha(F_l)$ means the area of $\alpha$ shape of points related to the $F_l$, $A(F_l)$ is the area of $F_l$, and the value of $A_\alpha(F_l)/A(F_l)$ can represent the coverage rate. This energy term represents the consistency between original point cloud and
 boundary faces of convex polyhedrons directly. 
The optimization is implemented by a max-flow algorithm that has been used in KSR~\cite{bauchet2020kinetic}. The corresponding binary labels of $\{q\}$ can be obtained when the direction energy is minimized. An instance is shown in Fig.~\ref{f5_winpoly}. In each iteration of the optimization strategy, the orientation of the candidate faces located at the boundary can be determined. Such faces are used to calculate $E_{dir}$ that labels the remainder polyhedrons in the next iteration. More details are described in the supplementary material.  

\begin{figure}[t]
  \centering
\includegraphics[width=0.7\linewidth]{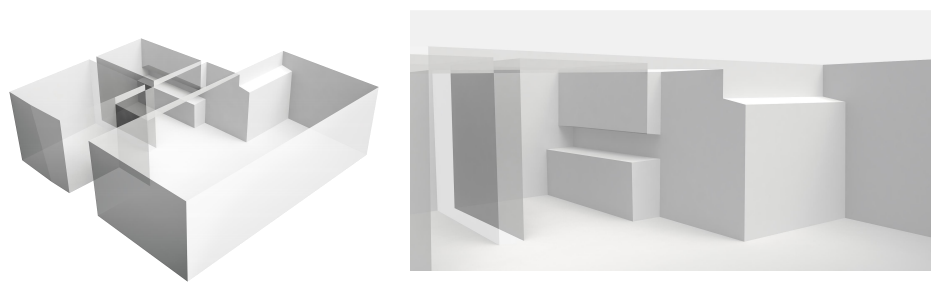} 
  \caption{Visualization of internal structure reconstructed by WindPoly.}  
  \label{f6_Inter}
\end{figure}

Benefiting from the accurate determined directions for polyhedron set, a more accurate internal structure (Fig.~\ref{f6_Inter}) can be checked and processed by the polygon-based winding numbers optimization. Related centroids involved in the optimization correspond to the number of polyhedrons, which is much less than the original points. Compared to the traditional winding numbers optimization, our method significantly improves efficiency.
\section{Experiment}\label{sec:results}

We evaluate the performance of WindPoly in the polygonal mesh reconstruction task. The experimental machine is equipped with Intel i9-13900K, 128GB RAM, RTX4090, with Windows 10 as the operation system and Visual Studio 2019 as the development platform. The experiments include the following parts: 1. we introduce the test dataset and explain some selected metrics to prove quantitative analysis for the reconstruction; 2. we compare different reconstruction methods to report and visualize the advantages of WindPoly; 3. we discuss some potential limitations of WindPoly in practice.

\begin{figure}[b]
	\centering 
    \includegraphics[width=\linewidth]{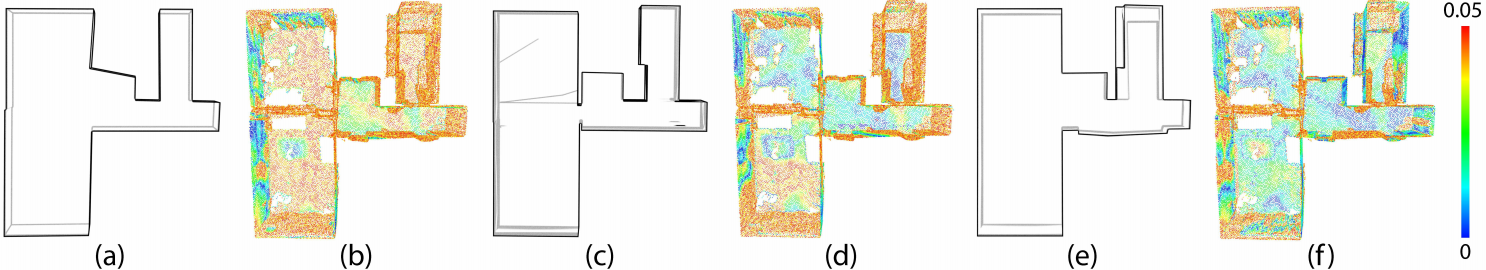} 
	\caption{Reconstruction results and color maps of mapping distances by different methods. (a, b) PolyFit, (c, d) KSR, (e, f) WindPoly.}  
	\label{fe1}
\end{figure}

\subsection{Dataset \& Metrics}

The test point clouds are collected from ABC dataset~\cite{Koch_2019_CVPR}, PolyFit dataset~\cite{nan2017polyfit}, UrbanBIS~\cite{UrbanBIS}, and BuildingNet~\cite{BuildingNet2021}, which reflect the reconstructed geometric details of different levels. The ABC dataset includes small-scale workpiece models with regular point distributions and clear geometric details. The PolyFit and BuildingNet datasets contain architectural point clouds. The UrbanBIS is a large scale city scene dataset with real scanned architectural point clouds. Based on the data scale mentioned in~\cite{gao2022low}, we select 300 models (100 models from ABC dataset and 200 models from other architectural datasets) with representative holes or smooth surfaces to evaluate the performance of the reconstruction for structured representation of geometric information.

The quantitative metrics should be established from two perspectives: geometric consistency and data compression efficiency. The geometric consistency can be represented by Hausdorff distance and mean distance, which characterizes the matching degree from points to the polygonal plane. For data compression efficiency, we directly report the point and face numbers of the reconstructed polygonal mesh. It should be noticed that the quality of the data compression should related to the geometric consistency. If the reconstruction result just contains one point, the better performance of data compression means nothing. Therefore, we multiple the Hausdorff distance and simplification rate to be a multiple estimation for fair evaluation.

\subsection{Comparisons}

Based on the collected dataset and related metrics, we evaluate the performance of different reconstruction methods, including PolyFit~\cite{nan2017polyfit}, KSR~\cite{bauchet2020kinetic}, IPSR~\cite{hou2022iterative}, LowPoly~\cite{gao2022low}, and R-LowPoly~\cite{chen2023robust}. Such methods cover the current mainstream solutions. However, the LowPoly and R-LowPoly can not reconstruct mesh from a raw point cloud directly. We employ a voxel-based Delaunay triangulation method~\cite{lv2021voxel} to achieve an initial mesh at first. The IPSR achieves a high-quality mesh without data compression. To provide a fair comparison, we employ the simplification from VCG library to concise the mesh.

\begin{figure}[t]
	\centering
    \includegraphics[width=\linewidth]{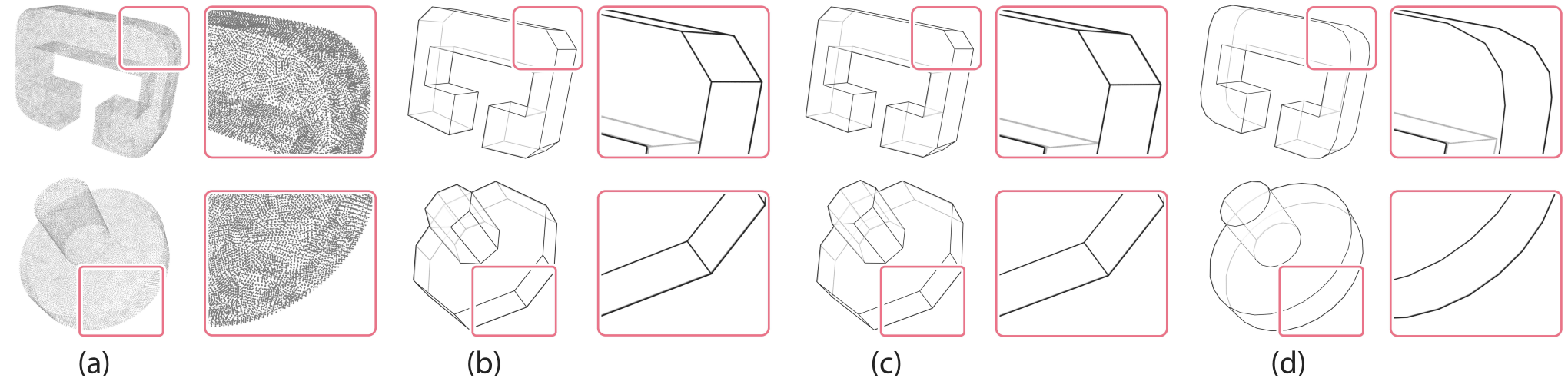} 
	\caption{External geometric details representation by different methods. (a) CAD model, (b) PolyFit, (c) KSR, (d) WindPoly.}  
	\label{fe1_1}
\end{figure}

\begin{figure}[t]
	\centering
    \includegraphics[width=\linewidth]{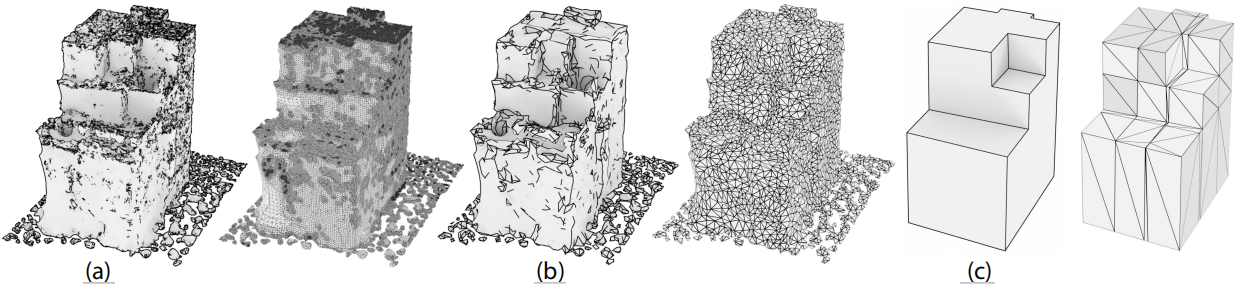}
	\caption{Visualization of reconstructed meshes and related triangulations by IPSR and WindPoly. (a) IPSR, point = 199.1k, face = 398.1k, (b) IPSR with simplification, point = 5.0k, face = 9.9k, (C) WindPoly, point = 336, face = 168. WindPoly achieves better structural planes and sharp features.}  
	\label{fe2}
\end{figure}

To show the performance of internal geometric structure reconstruction, we compute color maps for reconstructed meshes using different low-poly meshing methods. Mapping distances from points to related planes are used to generate colors. The upper layer of the point cloud is peeled off to display the internal color differences. Such color maps are shown in Fig.~\ref{fe1}. Comparing to other low-poly meshing schemes, the fitting errors of WindPoly in the internal geometric structure are lower. As mentioned before, WindPoly achieves a balance between planes with different scales. It keeps more accurate geometric features in regions with smooth curvature transitions. In Fig.~\ref{fe1_1}, we compare some results by classical low-poly meshing methods and WindPoly. It is clear that our method achieves more accurate results with better geometric consistency. Related qualitative results are reported in Tables~\ref{T1_qa} and~\ref{T2_qa}.

\begin{figure}[t]
	\centering
\includegraphics[width=\linewidth]{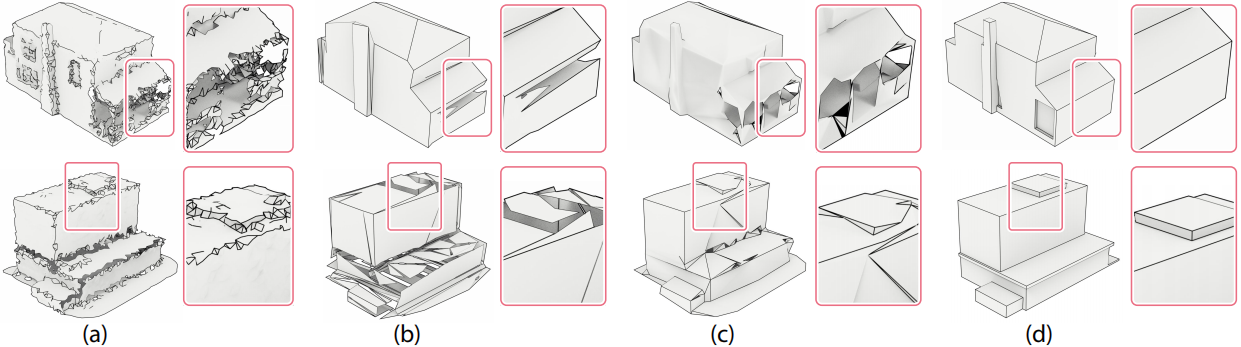} 
	\caption{Visualization of reconstructed meshes by different methods. (a) Initial mesh by \cite{lv2021voxel}, (b) LowPoly, (C) R-LowPoly , (d) WindPoly. }  
	\label{fe3_1}
\end{figure}

\begin{table}[t]
\centering
\caption{Quantitative analysis of different structural reconstruction methods in CAD models from ABC dataset. The related metrics include Hausdorff distance $Dis^H$, mean distance $Dis^M$, average point number $p^{Avg.}$, average face number $F^{Avg.}$, simplification rate $R^{Avg.}$, and multiple estimation $RH^{Avg.}=Dis^H\times R^{Avg.}$.}
\label{T1_qa}
 \setlength{\tabcolsep}{6pt}    
\resizebox{0.8\columnwidth}{!}{
\begin{tabular}{l|cccccc}
\toprule 
\textbf{}              & \textbf{Dis$^{H}\downarrow$} & \textbf{Dis$^{M}\downarrow$} & \textbf{p$^{Avg.}\downarrow$} & \textbf{F$^{Avg.}\downarrow$} & \textbf{R$^{Avg.}\downarrow$} & \textbf{RH$^{Avg.}\downarrow$} \\ \midrule
\textbf{PolyFit~\cite{nan2017polyfit}}   & 0.118    & 0.014   & 277  & 299   & 0.017    & 0.0011   \\                                                
\textbf{KSR~\cite{bauchet2020kinetic}}    & 0.111    & 0.015   & \textbf{61}  & 123   & 0.005    & 0.0005   \\                                  
\textbf{LowPoly~\cite{gao2022low}}    & 0.281    & 0.018   & 62  & \textbf{114}   & \textbf{0.005}    & 0.0011   \\              
\textbf{R-LowPoly~\cite{chen2023robust}}   & 0.267    & 0.014   & 246  & 368   & 0.009    & 0.0019   \\      
                                           
\textbf{WindPoly}   & \textbf{0.061}    & \textbf{0.002}   & 273  & 447   & 0.024    & \textbf{0.0010}   \\
\bottomrule                                    
\end{tabular}}
\end{table}

\begin{figure}[t]
  \centering
\includegraphics[width=\linewidth]{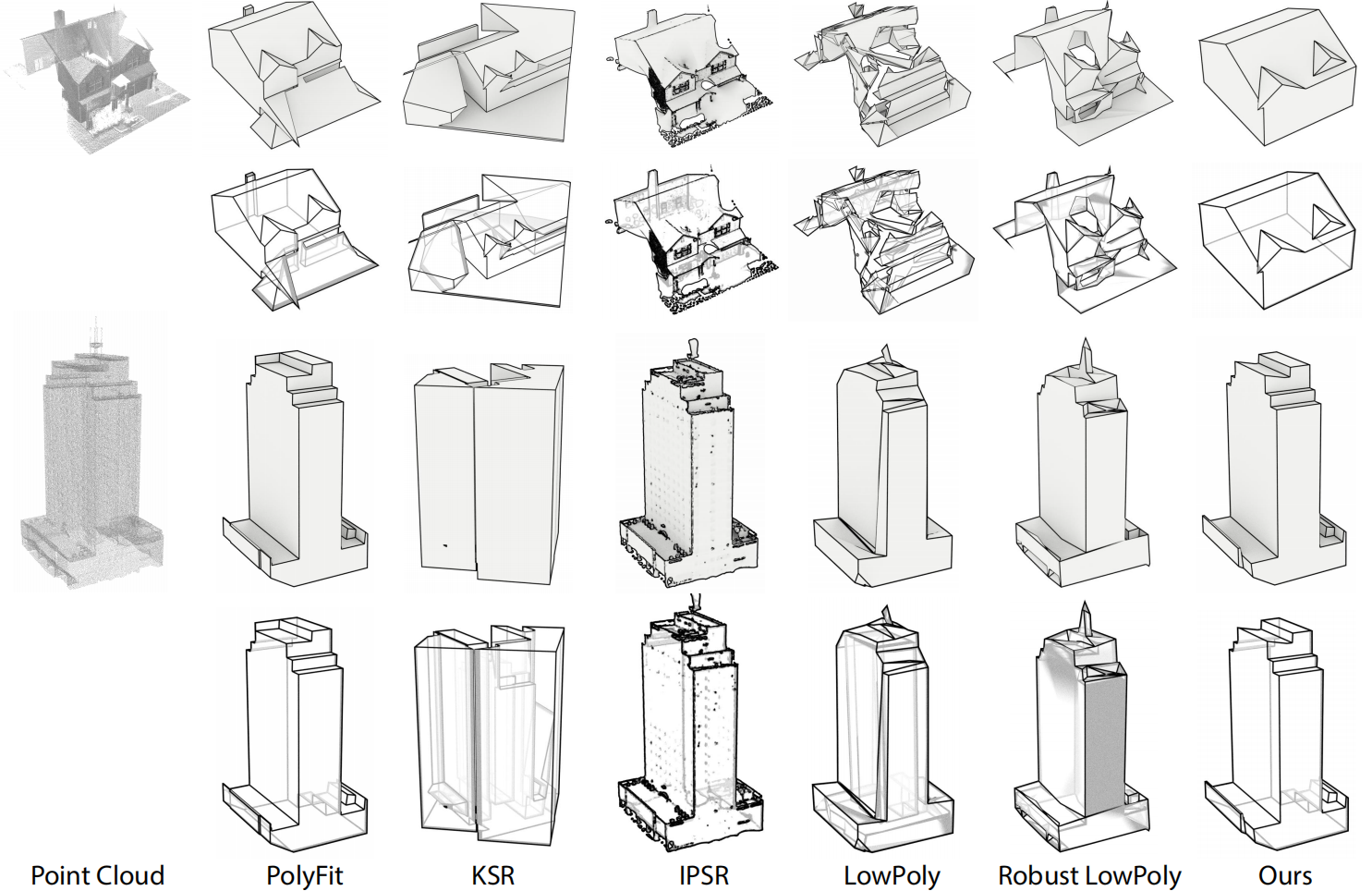} 
  \caption{Comparisons of different polygonal mesh reconstruction methods.}  
  \label{fe3}
\end{figure}

\begin{table}[t]
\centering
\caption{Quantitative analysis of different structural reconstruction methods in building models from architectural dataset.}
\label{T2_qa}
\setlength{\tabcolsep}{6pt}    
\resizebox{0.8\columnwidth}{!}{
\begin{tabular}{l|cccccc}
\toprule 
\textbf{}              & \textbf{Dis$^{H}\downarrow$} & \textbf{Dis$^{M}\downarrow$} & \textbf{p$^{Avg.}\downarrow$} & \textbf{F$^{Avg.}\downarrow$} & \textbf{R$^{Avg.}\downarrow$} & \textbf{RH$^{Avg.}\downarrow$} \\ \midrule
\textbf{PolyFit~\cite{nan2017polyfit}}    & 5.1652&	0.7808&	392&	433&	0.00607&	0.0496   \\                                                
\textbf{KSR~\cite{bauchet2020kinetic}}   & 6.7792	&0.6845&	\textbf{157}&	312&	\textbf{0.00362}&	0.0398  \\  

\textbf{LowPoly~\cite{gao2022low}}    & 6.6763&	\textbf{0.4863}&	378&	\textbf{188}&	0.00471&	0.0313  \\  

\textbf{R-LowPoly~\cite{chen2023robust}}    & 6.6121&	1.8565&	476&	1064&	0.01181&	0.0283   \\    
                                              
\textbf{WindPoly}    & \textbf{4.5276} & 	0.5159 & 	549 & 	274 & 	0.00560 & 	\textbf{0.0229}   \\
\bottomrule                                    
\end{tabular}}
\end{table}

To further compare the difference between the two-step reconstruction (mesh reconstruction with simplification) and WindPoly, we exhibit an independent comparison between IPSR and WindPoly. The rendering results of reconstructed meshes are shown in Fig.~\ref{fe2}. Benefited from the accurate implicit surface estimation, IPSR achieves more precise geometric details. However, the reconstructed mesh by IPSR take more points and faces even it has been simplified. Some sharp features are smoothed. In contrast, the performance of WindPoly for data compression is significantly improved while keeping sharp features. More results are shown in Fig.~\ref{fe3}. As structural methods, LowPoly and R-LowPoly can establish polygonal models with concise structures. However, both methods rely on the quality of the initial mesh. For some broken regions as shown in Fig.~\ref{fe3_1}, such methods cannot automatically repair them and result in noticeable topological errors. In Tables~\ref{T1_qa} and~\ref{T2_qa}, more qualitative results of structural reconstruction methods are reported. WindPoly achieves better results without initial meshes. 

\begin{figure}[t]
  \centering
  \includegraphics[width=\linewidth]{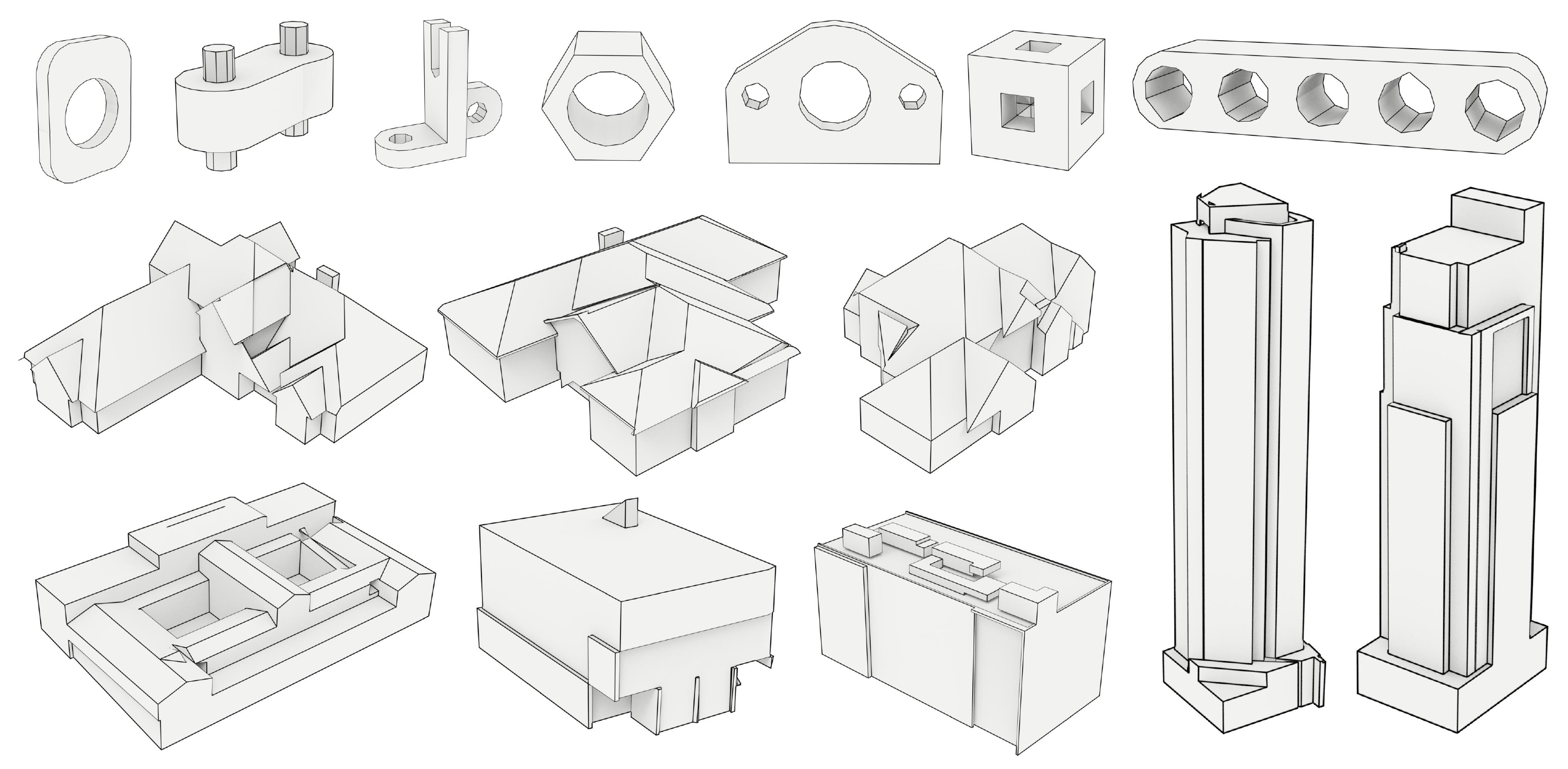} 
  \caption{More practical instances of polygonal meshes by WindPoly.}  
  \label{fe4}
\end{figure}

For efficiency analysis, we show time cost reports of different methods in Table~\ref{T3_time}. KSR is faster but it depends on normal vectors and may lose some geometric structures like instances in Fig.~\ref{fe3}. For complex architectural point clouds, IPSR takes more computational cost and outputs meshes with redundant points and faces. LowPoly and R-LowPoly require initial mesh reconstruction. In contrast, WindPoly achieves more concise representation and better geometric structure. It improves the performance for polyhedron searching by the spatial partition strategy which avoids low-contributing planes in convex polyhedron generation. The polygon-based winding numbers optimization further improves the computational efficiency. In summary, WindPoly provides a balanced and effective solution for polygonal mesh reconstruction, as shown in Fig.~\ref{fe4}.

\textbf{Limitations.} WindPoly is effective for point clouds with significantly structural information. However, once such information of the point cloud is not the main content, the performance of WindPoly is affected to some extent. For some smaller structures in large-scale point clouds, there is a certain probability of producing incorrect results, especially when the point density of the structure is lower. Such smaller structures may be merged into incorrect planes. Some instances are shown in supplementary materials.

\begin{table}[t]
    \caption{
        Time cost reports of different methods in related stages. Initial pre-processing of PolyFit and WindPoly is polyhedron searching; Initial mesh reconstruction of LowPoly and R-LowPoly is voxel-based Delaunay triangulation method~\cite{lv2021voxel}.
    }
    \label{T3_time}
    \centering
    \setlength{\tabcolsep}{3pt}  
    \resizebox{\columnwidth}{!}{
    \begin{tabular}{l|cccccccccccccccccc}
        \toprule
        \textbf{Method} && \multicolumn{2}{c}{\textbf{PolyFit}}  && \multicolumn{2}{c}{\textbf{KSR}}&& \multicolumn{2}{c}{\textbf{IPSR}}&& \multicolumn{2}{c}{\textbf{LowPoly}}&& \multicolumn{2}{c}{\textbf{R-LowPoly}}&& \multicolumn{2}{c}{\textbf{WindPoly}}\\
        \cmidrule(lr){3-4} \cmidrule(lr){6-7} \cmidrule(lr){9-10} \cmidrule(lr){12-13} \cmidrule(lr){15-16} \cmidrule(lr){18-19}
        && Initial & Total && Initial & Total && Initial & Total && Initial & Total&& Initial & Total&& Initial & Total \\
        \midrule
        \textbf{ABC Dataset} && 5s & 10s && — & 6s && —& 16s && 17s & 58s && 17s & 78s && 31s & 133s  \\
        \textbf{PolyFit} && 6s & 15s && — & 9s && — & 345s && 11s & 103s && 11s & 120s && 19s & 99s \\ 
        \textbf{BuildingNet} && 56s & 803s && — & 7s && — & 239s && 10s & 80s && 10s & 78s && 20s & 76s  \\   
        \textbf{UrbanBIS} && 330s & 1707s && — & 11s && — & 1,347s && 9s & 196s && 9s & 98s && 25s & 204s  \\   
        \bottomrule
    \end{tabular}}
\end{table} 
\section{Conclusion}

We propose a polygonal mesh reconstruction method WindPoly to reconstruct a concise 3D representation with accurate structural information for raw point clouds. Three core parts work together to achieve this goal. The polygonal plane detection implements the initial processing to extract primitive planes from the point cloud. Based on the planes, the adaptive spatial partition checks the convex polyhedral space and generate a set of polyhedrons without orientation. Some internal geometric structures are preserved. Finally, the polygon-based winding numbers optimization orients the faces of the polyhedrons and outputs the reconstructed polygonal mesh. The WindPoly can establish structural information from raw point clouds without point-based normal vector assistance. It orients the polygonal mesh with an efficient way while capturing the accurate internal geometric structures. Experiments show that WindPoly achieves a better balance between normal vector independence, geometric consistency and data compression. It can handle CAD models and  architectural point clouds, and output high quality polygonal meshes.

\section*{Acknowledgements}
This work was supported in parts by NSFC (U21B2023, U2001206, 62161146005), Guangdong Basic and Applied Basic Research Foundation (2023B1515120026, 2023A1515110292), DEGP Innovation Team (2022KCXTD025), Shenzhen Science and Technology Program (KQTD20210811090044003, RCJC2020071411443 5012, JCYJ20210324120213036), Development Funds from Shenzhen University and Guangdong Laboratory of Artificial Intelligence and Digital Economy (SZ).

%
%
\bibliographystyle{splncs04}

\end{document}